\documentclass[reprint,aps,prl]{revtex4-2}
\usepackage{graphicx}% Include figure files
\usepackage{bm}% bold math %\bibliographystyle{apsrev}
\usepackage{bbold}
\usepackage{CJK}
\usepackage{amsthm}
\usepackage{multirow}
\usepackage{xcolor}
\usepackage{MnSymbol}%
\usepackage{wasysym}%
\usepackage{threeparttable}

\usepackage{amsmath}

\theoremstyle{plain}
\newtheorem*{theorem*}{Theorem}

\begin{document}

\title{Multipartite Entanglement: A Journey Through Geometry}

\author{Songbo Xie}
\email{sxie9@ur.rochester.edu}
\affiliation{Center for Coherence and Quantum Optics, and Department of Physics and Astronomy, University of Rochester, Rochester, New York 14627, USA}

\author{Daniel Younis}
\affiliation{Center for Coherence and Quantum Optics, and Department of Physics and Astronomy, University of Rochester, Rochester, New York 14627, USA}

\author{Yuhan Mei}
\affiliation{Center for Coherence and Quantum Optics, and Department of Physics and Astronomy, University of Rochester, Rochester, New York 14627, USA}

\author{Joseph H. Eberly}
\affiliation{Center for Coherence and Quantum Optics, and Department of Physics and Astronomy, University of Rochester, Rochester, New York 14627, USA}

\date{\today}

\begin{abstract}
Genuine multipartite entanglement is crucial for quantum information and related technologies but quantifying it has been a long-standing challenge. Most proposed measures do not meet the ``genuine'' requirement, making them unsuitable for many applications. In this work, we propose a journey toward addressing this issue by introducing an unexpected relation between multipartite entanglement and hypervolume of geometric simplices, leading to a tetrahedron measure of quadripartite entanglement. By comparing the entanglement ranking of two highly entangled four-qubit states, we show that the tetrahedron measure relies on the degree of permutation invariance among parties within the quantum system. We demonstrate potential future applications of our measure in the context of quantum information scrambling within many-body systems.
\end{abstract}

\maketitle

{\it Introduction}.---Genuine multipartite entanglement (GME) is a form of entanglement beyond two parties, offering distinct advantages over two-party entanglement \cite{dur2000}. The primary benefit of GME lies in its capacity to facilitate complex quantum tasks \cite{laflamme1996,chin2012,karlsson1998}. Consequently, comprehending GME in many-body systems is valuable for the development of quantum technologies.

An entanglement measure is necessary to quantify the amount of entanglement. Entanglement measures are vital in various actions, including investigating the dynamical behavior of many-body systems, quantifying the performance of quantum protocols, and identifying the nature of phase transitions in many-body systems \cite{ding2021,anshu2022,block2022,puliyil2022}. The ``genuine'' requirement requires that a GME measure should only, and always, vanish for biseparable states \cite{ma2011,xie2021}. This is crucial to quantify accurately the collective strength of the parties' entanglement and unlock their potential in quantum tasks.

Unfortunately, many proposed multipartite entanglement measures, including some extensively used \cite{coffman2000,meyer2002,carvalho2004,jungnitsch2011}, fail to meet the ``genuine'' requirement, making them unsuitable in many practical applications. One popular route to construct a GME measure is to use distance-based approaches, including relative entropy \cite{vedral1998entanglement} and some ``geometric measures'' \cite{wei2003geometric,blasone2008hierarchies,sen2010channel}. However, evaluating such measures requires using optimization procedures, which is highly challenging, sometimes even for pure states. 

In this work, to solve this measure-identification problem, we embark on a distinct journey. The journey begins with the triangle measure for three-qubit systems that we introduced recently \cite{xie2021}. However, extending this measure to systems with more than three qubits is challenging due to the over-flexibility of geometric shapes. Various attempts have been made, but they either fail to meet the ``genuine'' requirement or lose simple geometric intuition \cite{mishra2022, guo2022,jin2023}. 

We demonstrate the possibility of a ``geometric journey'' by presenting a tetrahedron measure for four-qubit systems. We show that our novel method, even though sharing the same term ``geometry'' with distance-based measures, presents fundamental distinctions by offering an unexpected connection between multipartite entanglement and hypervolume of geometric simplices. It also relies on the degree of permutation invariance among parties within the quantum system---a feature not shared by the distance-based measures. This promises future avenues, for example, for investigating information scrambling in many-body systems.

{\it From three to four}.---Von Neumann entropy is crucial in our approach. Specifically, the entanglement $E$ of a bipartite pure quantum state $\psi_{12}$ can be characterized by the von Neumann entropy $S$ of either of the individual parties, $E(\psi_{12}):=S(\rho_i)\equiv-\text{Tr}\left(\rho_i\log_2\rho_i\right)$, where $\rho_i$ is the reduced density operator of the party $i$. $E(\psi_{12})$ is called {\it entanglement entropy} of the state $\psi_{12}$.

Other crucial ingredients are the ``entanglement polygon inequalities'' \cite{qian2018}. Specifically, for a general $n$-qubit system, when entanglement entropy is considered between one qubit and the remaining $(n-1)$ qubits, we obtain $n$ such one-to-other bipartite entropies. The entanglement polygon inequalities require that any one of them cannot exceed the sum of the other $(n-1)$ values. For a three-qubit state, these relations hold: $E_{i(jk)}\leq E_{j(ki)}+E_{k(ij)}$. Here, $E_{i(jk)}:= S(\rho_i)$, where the subscript $i$ refers to the system's $i$th qubit, and $\{i,j,k\}$ is any permutation of $\{1,2,3\}$. 

Those inequalities can be geometrically interpreted as the three one-to-other entropies representing the lengths of the three edges of a triangle. We demonstrated that the area of this triangle is a three-qubit GME measure, with concurrence being employed in lieu of entanglement entropy \cite{xie2021,xie2022}. We now illustrate that entanglement entropy offers distinct advantages.

The inequalities for four-qubit systems are \cite{qian2018}:
\begin{equation}\label{4qubit}
    E_{i(jkl)}\leq E_{j(kli)}+E_{k(lij)}+E_{l(ijk)},
\end{equation}
with $\{i,j,k,l\}$ being any permutation of $\{1,2,3,4\}$. As suggested in \cite{qian2018}, Eq.~\eqref{4qubit} induces a quadrilateral, with the four edge lengths equal to the four one-to-other entropies. Therefore, following from the triangle measure in \cite{xie2021}, a tempting candidate of a four-qubit GME measure is the area of this quadrilateral.

\begin{figure}[t]
    \centering
    \includegraphics[width=0.45\textwidth]{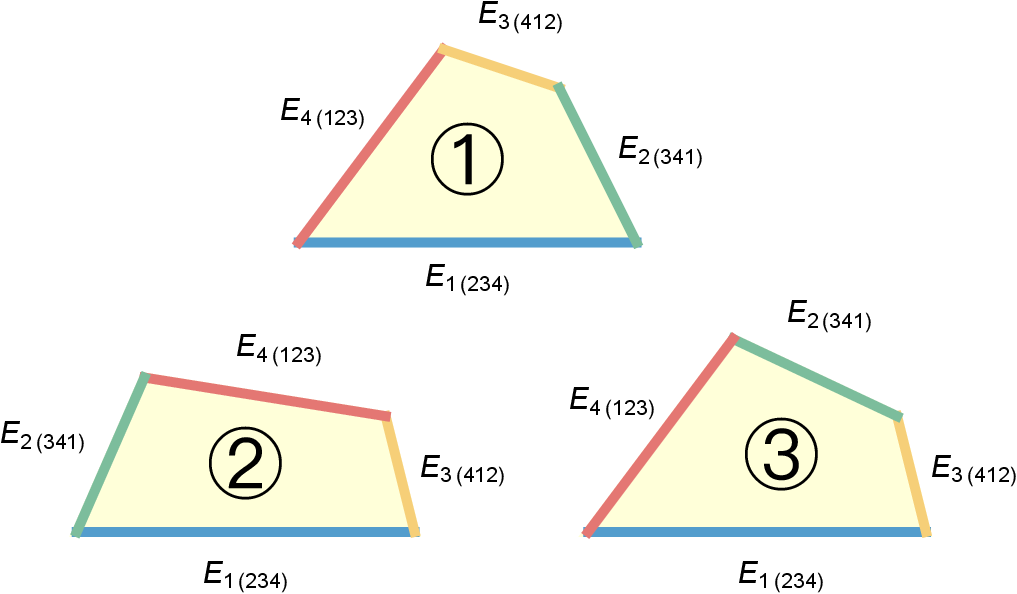}
    \caption{The interpretation of Eq.~\eqref{4qubit} as a quadrilateral violates the invariance of entanglement under qubit permutations since the four edges are not treated equally, resulting in three distinct possible quadrilaterals. Here, edges of the same color indicate equal length.}
    \label{fig:polygonsimplex}
\end{figure}

However, using the quadrilateral defined above poses problems. Firstly, since entanglement is invariant under qubit permutations, the four one-to-other entropies $E_{i(jkl)}$ must be treated equally. However, the four edges of a quadrilateral are not necessarily equal. A quadrilateral with different edge lengths can be arranged in three different ways, as shown in Fig.~\ref{fig:polygonsimplex}. Furthermore, a quadrilateral has five degrees of freedom, while the edge lengths fix only four of them. Consequently, even if the four edge lengths are fixed, the area of the quadrilateral can still vary, rendering the entanglement measure undetermined.

{\it From polygon to simplex}.---Instead, we consider an ``entropic tetrahedron'' induced by Eq.~\eqref{4qubit}, of which the four face areas are equal to the four one-to-other entropies, as shown in Fig.~\ref{fig:polygonsimplex}. The advantage of this tetrahedron approach quickly becomes apparent when one observes that the four faces are treated equally. We reinterpret the results of \cite{qian2018} in this new framework as {\it entanglement simplex inequalities}, where the term ``simplex'' generalizes the notions of triangle and tetrahedron to arbitrary dimensions.

Before proceeding, it is important to note that a tetrahedron has six degrees of freedom, while the four face areas fix only four of them. Consequently, if we attempt to employ the entropic tetrahedron's volume to measure GME, it remains undetermined.

\begin{figure}[t]
    \centering
    \includegraphics[width=0.35\textwidth]{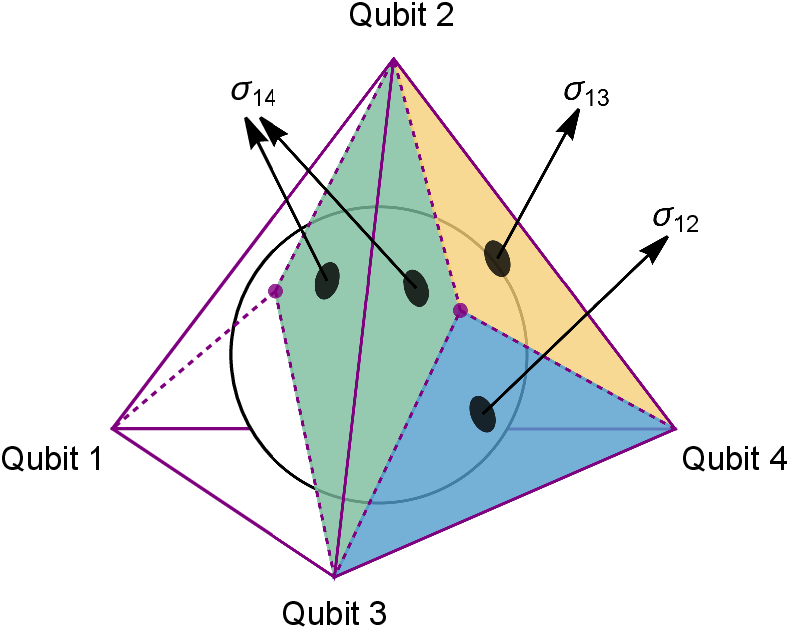}
    \caption{The inscribed sphere for a tetrahedron divides the four faces into twelve smaller triangles. Any two smaller triangles that share an edge of the tetrahedron exhibit identical areas, represented by the same color.}
    \label{fig:sphere}
\end{figure}

To further elaborate, we note that every tetrahedron has a unique inscribed sphere tangent to all four faces. For each face, one can connect the point of contact with the inscribed sphere to the three vertices, which divides the face into three smaller triangles. The twelve smaller triangles on the four faces are related pairwise: any two triangles that share an edge of the tetrahedron have the same area. The six different areas of the smaller triangles are labeled $\{\sigma_{12},\sigma_{13},\sigma_{14},\sigma_{23},\sigma_{24},\sigma_{34}\}$, shown in Fig.~\ref{fig:sphere}. As a simple exercise, one can show that the areas of the smaller triangles fix the tetrahedron, whose volume is:
\begin{equation}\label{volume}
    \begin{split}
        V&=\dfrac{\sqrt{2}}{3}S^{1/2}(A_0A_1A_2A_3)^{1/4},\\
    \text{with}\     A_0&=+\sqrt{\sigma_{12}\sigma_{34}}+\sqrt{\sigma_{13}\sigma_{24}}+\sqrt{\sigma_{14}\sigma_{23}},\\
        A_1&=-\sqrt{\sigma_{12}\sigma_{34}}+\sqrt{\sigma_{13}\sigma_{24}}+\sqrt{\sigma_{14}\sigma_{23}},\\
        A_2&=+\sqrt{\sigma_{12}\sigma_{34}}-\sqrt{\sigma_{13}\sigma_{24}}+\sqrt{\sigma_{14}\sigma_{23}},\\
        A_3&=+\sqrt{\sigma_{12}\sigma_{34}}+\sqrt{\sigma_{13}\sigma_{24}}-\sqrt{\sigma_{14}\sigma_{23}}.
\end{split}
\end{equation}
and $S=2(\sigma_{12}+\sigma_{13}+\sigma_{14}+\sigma_{23}+\sigma_{24}+\sigma_{34})$, the total surface area of the entropic tetrahedron. 

The six parameters $\{\sigma_{ij}\}$ must add up to the four face areas induced by the ``entanglement simplex inequalities'', given by (including four index permutations):
\begin{equation}\label{eqn1}
    \sigma_{ij}+\sigma_{ik}+\sigma_{il}=E_{i(jkl)}.
\end{equation}
However, these equations cannot fix the tetrahedron, as two more equations are required. While the one-to-other entropies have been used so far, the construction of a four-qubit GME must also involve entropies representing two-to-other bipartitions, namely, $\{E_{(12)(34)},E_{(13)(24)},E_{(14)(23)}\}$, with $E_{(ij)(kl)}:=S(\rho_{ij})\equiv S(\rho_{kl})$, where the bipartitions separate the qubits into two groups of two.

On the one hand, if any one of the three two-to-other entanglement entropies vanishes, the state is biseparable. According to the ``genuine'' requirement, our proposed GME measure, the volume of the tetrahedron, should be zero. On the other hand, if any one of the three quantities $A_1$, $A_2$, and $A_3$ in Eq.~\eqref{volume} vanishes, the tetrahedron's volume is also zero. To ensure consistency between these two conditions, we relate the quantities by the following equations (including three index permutations):
\begin{equation}\label{eqn2}
    \begin{split}
        -\sqrt{\sigma_{ij}\sigma_{kl}}+\sqrt{\sigma_{ik}\sigma_{jl}}+\sqrt{\sigma_{il}\sigma_{jk}}=\lambda\cdot E_{(ij)(kl)}.
    \end{split}
\end{equation}
The additional parameter $\lambda$ can be canceled out when dividing the equations in \eqref{eqn2} in pairs, leading to six equations for $\{\sigma_{ij}\}$ by combining Eq.~\eqref{eqn1} and \eqref{eqn2}.

{\it Existence of solutions}.---Eq.~\eqref{eqn2} can be transformed to the quadratic form: $\sigma_{ij}\sigma_{kl}=(\lambda^2/4)[E_{(ik)(jl)}+E_{(il)(jk)}]^2$. Together with Eq.~\eqref{eqn1}, this simplifies the equations into one quartic equation for one single variable $\sigma_{ij}$ with four distinct solutions. Importantly, this reasoning is universally applicable to all six $\{\sigma_{ij}\}$ due to their inherent symmetry. Consequently, a total of four sets of solutions exist for the six equations.

Our numerical investigations consistently confirm the presence of four distinct solution sets. Within these sets, there is one unique solution where all $\{\sigma_{ij}\}$ have nonnegative values. Consequently, given an arbitrary four-qubit state, a tetrahedron can always be constructed where the areas of the six smaller triangles of the tetrahedron are equal to the nonnegative values $\{\sigma_{ij}\}$. We call the construction an {\it entropic tetrahedron}.

We introduce the four-qubit {\it entropic fill}, denoted as $F_4:=(3^{7/6}/2)V^{2/3}$, with $V$ being the tetrahedron volume, given in Eq.~\eqref{volume}. The prefactor $3^{7/6}/2$ ensures the normalization $0\leq F_4\leq1$. As will be explained, the exponent $2/3$ contributes to additivity of $F_4$. We now prove that $F_4$ is a GME measure for four-qubit pure states.

(\textit{A}){\it $F_4$ satisfies the ``genuine'' requirement}.---For a four-qubit pure system, there are two types of biseparable states: one-to-other biseparable and two-to-other biseparable. In the case of one-to-other biseparable states, all four quantities $A_0$, $A_1$, $A_2$, and $A_4$ are zero, resulting in a tetrahedron with zero volume. For two-to-other biseparable states, one of the three quantities $A_1$, $A_2$, or $A_3$ is zero, leading to a tetrahedron with zero volume as well.

Conversely, if the volume of the constructed tetrahedron is zero, there are four possible scenarios, based on the expression of $V$ in Eq.~\eqref{volume}:\\
(a) $(A_1A_2A_3)=0$, but $A_1$, $A_2$, and $A_3$ are not all zero. Without loss of generality, we assume $A_1=0$. According to Eq.~\eqref{eqn2}, this leads to $\lambda\neq0$ and $E_{(12)(34)}=0$, indicating that the state is at least two-to-other biseparable.\\
(b) $A_0=0$. This is equivalent to $A_1=A_2=A_3=0$. In this scenario, it is easy to observe that $\sqrt{\sigma_{12}\sigma_{34}}=\sqrt{\sigma_{13}\sigma_{24}}=\sqrt{\sigma_{14}\sigma_{23}}=0$. Therefore, at least one factor under each square root should be chosen to equal zero. We first assume that there exists a subscript $i$ such that $\sigma_{ij}=\sigma_{ik}=\sigma_{il}=0$. This leads to $E_{i(jkl)}=0$, indicating that the state is at least one-to-other biseparable.\\
(c) $A_0=0$, but the assumption in (b) is not valid. It can be shown that the only possibility to maintain each square root being zero is $\sigma_{jk}=\sigma_{kl}=\sigma_{lj}=0$, while $\sigma_{ij}\neq0$, $\sigma_{ik}\neq0$, and $\sigma_{il}\neq0$. The four face areas satisfy the relations $E_{i(jkl)}=E_{j(kli)}+E_{k(lij)}+E_{l(ijk)}$ with $E_{i(jkl)}>E_{j(kli)}>0$, $E_{i(jkl)}>E_{k(lij)}>0$, and $E_{i(jkl)}>E_{l(ijk)}>0$, representing a tetrahedron with its four vertices coplanar. Using the same approach as in the appendix of \cite{xie2021}, a contradiction can be shown. Therefore, this scenario does not exist for four-qubit systems.\\
(d) $S=0$. This implies that all $\{\sigma_{ij}\}$ are equal to zero. According to Eq.~\eqref{eqn1}, the state is a product state.

Therefore, we have demonstrated that the volume of the entropic tetrahedron assigned to a four-qubit state is zero if and only if the state is biseparable, demonstrating that $F_4$ satisfies the genuine requirement.

(\textit{B}){\it $F_4$ is additive}.---An entanglement measure $E$ is additive if $E(\psi^{\otimes n})=nE(\psi)$, indicating that the entanglement of $n$ identical copies of a state is $n$ times the entanglement of a single copy \cite{bennett1996mixed}. 

With von Neumann entropy's additivity being a known fact, the quantities $\{E_{i(jkl)},E_{(ij)(kl)}\}$ are all additive in Eq.~\eqref{eqn1} and \eqref{eqn2}. Consequently, for $n$ identical copies, the nonnegative solutions of $\{\sigma_{ij}\}$ increase by a factor of $n$. This establishes the additivity of $F_4$ and justifies the essential inclusion of the exponent $2/3$ for $V$.

\begin{figure*}[t]
    \centering
    \includegraphics[width=\textwidth]{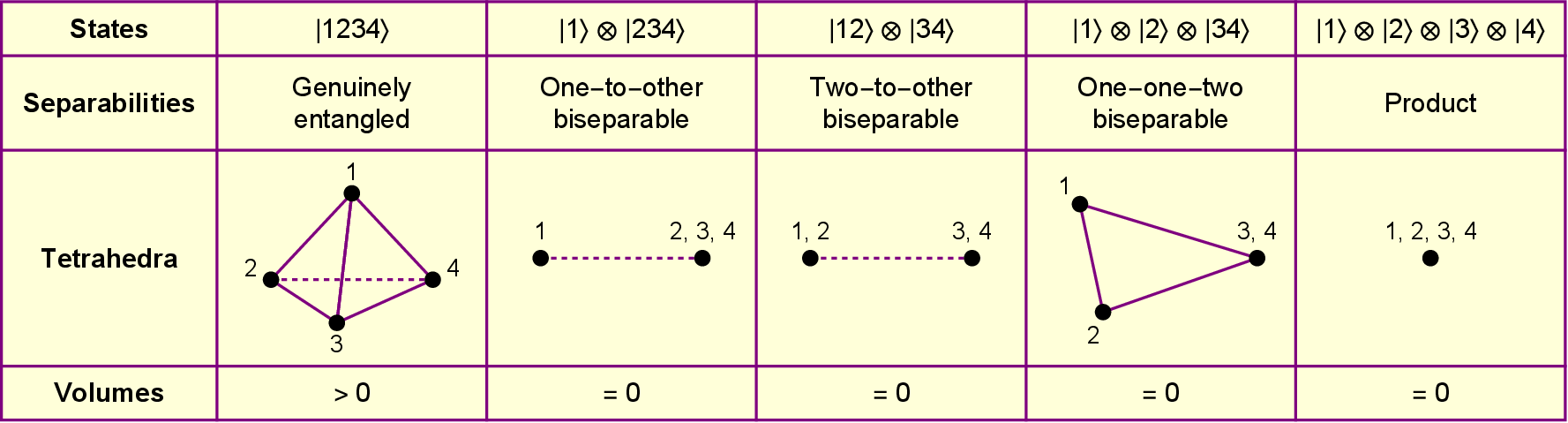}
    \caption{Entropic tetrahedra for different types of four-qubit states. (a) The tetrahedron for a genuinely entangled state has a positive volume. (b) The tetrahedron for a one-to-other biseparable state is an infinitely long line with zero volume, represented by the dashed line. (c) The tetrahedron for a two-to-other biseparable state is also an infinitely long line with zero volume, represented by the dashed line. (d) The tetrahedron for a one-one-two biseparable state is a coplanar triangle with zero volume. It is interesting to note that this is the only case where the shape of the tetrahedron is not fixed, as long as the area of the triangle is equal to the bipartite entropy $E_{34}$. (e) The tetrahedron for a product state is a dot with zero volume.}
    \label{fig:table}
\end{figure*}

(\textit{C}){\it Local monotonicity}.---An entanglement measure $E$ should also satisfy local monotonicity, requiring that $E$ is nonincreasing under local operations and classical communications (LOCC) \cite{vidal2000entanglement}. For a general mixed state $\rho$, it means $E(\rho)\geq E(\Lambda(\rho))$, where $\Lambda$ is an arbitrary LOCC process \cite{chitambar2014everything}. For measures only defined for pure states, a stronger form of the requirement applies: $E(\psi)\geq \sum_ip_iE(\varphi_i)$, where $\{p_i,|\varphi_i\rangle\}$ is the ensemble obtained from $|\psi\rangle$ under LOCC \cite{horodecki2009quantum}.

We verify local monotonicity of $F_4$ by adopting the numerical technique explained in \cite{ge2023tripartite}. In particular, as is shown in \cite{dur2000}, it suffices to confine LOCC operations to positive-operator-valued measures $\{X_1,X_2\}$ with binary outcomes, acting only on qubit 1. These operators satisfy the relation $X_1^\dagger X_1+X_2^\dagger X_2=\mathbb{1}$ and can be generally parameterized as:
\begin{equation}
    \begin{split}
        X_1&=\begin{pmatrix}
            \sin\alpha_1\cos\beta_1 & -e^{i\beta_2}\sin\alpha_1\sin\beta_1\\
            \sin\alpha_2\sin\beta_1 & e^{i\beta_2}\sin\alpha_2\cos\beta_1
        \end{pmatrix},\\
        X_2&=\begin{pmatrix}
            \cos\alpha_1\cos\beta_1 & -e^{i\beta_2}\cos\alpha_1\sin\beta_1\\
            \cos\alpha_2\sin\beta_1 & e^{i\beta_2}\cos\alpha_2\cos\beta_1
        \end{pmatrix},
    \end{split}
\end{equation}
with $\{\alpha_i,\beta_i\}\in[-\pi,\pi]$. Since entanglement measures are invariant under local unitary transformations, it suffices to confine the input state $\psi$ to the generalized Schmidt form \cite{acin2001three}:
\begin{equation}
    \begin{split}
        |\psi\rangle&=c_0|0000\rangle+c_4|0100\rangle+c_5|0101\rangle+c_6|0110\rangle\\
        &+c_8|1000\rangle+c_9e^{i\phi_1}|1001\rangle+c_{10}e^{i\phi_2}|1010\rangle\\
        &+c_{11}e^{i\phi_3}|1011\rangle+c_{12}e^{i\phi_4}|1100\rangle+c_{13}e^{i\phi_5}|1101\rangle\\
        &+c_{14}e^{i\phi_6}|1110\rangle+c_{15}e^{i\phi_7}|1111\rangle,
    \end{split}
\end{equation}
with $\sum_ic_i^2=1$ and $\{\phi_i\}\in[0,\pi]$. The binary outcome states are given by $|\varphi_i\rangle=(X_i\otimes\mathbb{1}\otimes\mathbb{1}\otimes\mathbb{1})|\psi\rangle/\sqrt{p_i}$, and $\{p_i\}$ are normalization factors. Local monotonicity of $F_4$ is equivalent to the inequality:
\begin{equation}
    M(c_i,\phi_i,\alpha_i,\beta_i):= \left[F_4(\psi)-\sum_i\big(p_iF_4(\varphi_i)\big)\right]\geq 0.
\end{equation}
We have conducted thorough numerical verifications, revealing that the minimum value of $M$ among the 22 parameters is on the order of $-10^{-6}$. We emphasize that the numerical uncertainty associated with $M$, stemming from numerically solving equations \eqref{eqn1} and \eqref{eqn2}, is of a comparable magnitude, with $\Delta M\sim10^{-6}$. To elaborate, we employ the propagation of uncertainty formula: $\Delta M=|\partial M/\partial \sigma_{ij}|\Delta\sigma_{ij}$, where $\Delta\sigma_{ij}\sim 10^{-15}$ arises from the double-precision format of $\sigma_{ij}$. Remarkably, our analysis indicates that $|\partial M/\partial \sigma_{ij}|$ can attain the order of $10^9$, confirming the overall uncertainty in $M$ as $\Delta M\sim10^{-6}$. Therefore, the minimum value of $M$ being $-10^{-6}$ only reflects the systematic uncertainties embedded in the numerical computations.

For further validation, we fix the parameters $\{c_i, \phi_i, \alpha_i, \beta_i\}$ that previously resulted in the negative values of $M$. We repeatedly solve the equations \eqref{eqn1} and \eqref{eqn2}. Remarkably, the computed value of $M$ exhibits oscillations, frequently transitioning to positive values across multiple trials. This observation illustrates that the minimum value of $F_4$ is effectively zero within the numerical-uncertainty tolerance. Consequently, this provides evidence that the entropic fill $F_4$ satisfies the stronger version of local monotonicity for four-qubit pure states. 

Now that we have confirmed that the quantity $F_4$ meets all criteria: (A) the ``genuine'' requirement, (B) additivity, and (C) local monotonicity, we claim that the entropic fill derived from the volume of the entropic tetrahedron is a proper four-qubit entanglement measure.

For a visual demonstration of entropic tetrahedra for different types of states, refer to Fig.~\ref{fig:table}. The caption provides detailed explanations. 

\begin{table}[b]
    \centering
    \begin{tabular}{c c c c c c}
          & $\{E_{i(jkl)}\}$ & $\{E_{(ij)(kl)}\}$ &  MIN & GM & $F_4$\\
         \hline\hline
       GHZ  & \{1,1,1,1\} & \{1,1,1\} & 1 & 1 & 1 \\
       \hline
       cluster  & \{1,1,1,1\} & \{1,2,2\} & 1 & 1.219 & 0.976\\
       \hline
    \end{tabular}
    \caption{Bipartite entanglement entropies, their minimum (labeled as MIN), their geometric mean (labeled as GM), and entropic fill $F_4$ for the GHZ state and the cluster state.}
    \label{tab:twostates}
\end{table}

{\it Geometry relies on symmetry}.---Entropic fill derives from the bipartite entropies $\{E_{i(jkl)}, E_{(ij)(kl)}\}$. We note that these values induce two additional entanglement measures through algebraic constructions: one being their minimum, labeled as ``MIN'', and the other being their geometric mean, labeled as ``GM.'' Constructions within these approaches can be found in \cite{ma2011, hashemi2012, schneeloch2020quantifying, li2022, ge2023tripartite}. Importantly, the two algebraic measures MIN and GM also satisfy the above three criteria (A), (B), and (C), establishing them as two additional entanglement measures. Next, we compare these three measures.

In four-qubit systems, the Greenberger-Horne-Zeilinger (GHZ) state $(|0000\rangle+|1111\rangle)/\sqrt{2}$ and the cluster state $|\phi_4\rangle=(|0000\rangle+|0011\rangle+|1100\rangle-|1111\rangle)/2$ are two famous examples of highly entangled states \cite{briegel2001}. These two states possess different entanglement properties and are used for different quantum tasks, so it is not straightforward to determine their entanglement ranking. Interestingly, we find that the above three measures give three entirely different entanglement rankings for the two states, as illustrated in Table~\ref{tab:twostates}.

Specifically, the MIN measure considers the two states as equally entangled, with $\text{MIN}(\text{GHZ})=1=\text{MIN}(\phi_4)$, as it only considers the smallest bipartite entropy, ignoring all the other details. The GM measure ranks the cluster state as more entangled, with $\text{GM}(\text{GHZ})=1<1.219=\text{GM}(\phi_4)$, as it multiplies all entropies together. 

The entropic fill measure presents a contrasting perspective by assigning a higher entanglement value to the GHZ state, evident in $F_4(\text{GHZ})=1>0.976=F_4(\phi_4)$. This distinction arises from its geometry-based construction, and geometry is tightly linked to symmetries. To illustrate, consider that greater symmetry corresponds to a larger volume---a principle demonstrated by the fact that a sphere is the most symmetric and has the maximal volume with a fixed total surface area. 

Applying this principle to our context, both the GHZ state and the cluster state tetrahedra share an identical total surface area. However, the tetrahedron associated with the GHZ state has a greater volume, signaling a higher degree of symmetry. This geometric symmetry comes from the permutation invariance among parties within the quantum system. Specifically, the GHZ state remains invariant under the permutation of all four qubits. In contrast, the cluster state exhibits permutation invariance only between the first two qubits and the last two qubits. Our tetrahedron measure of entanglement successfully captures that permutation invariance.

Furthermore, from our tetrahedron measure $F_4$, we can define the degree of permutation invariance $\mathcal{P}$ as follows: $\mathcal{P}:=F_4/F_{4,\text{max}}^{(S)}$, where $F_{4,\text{max}}^{(S)}$ is the maximal $F_4$ value with the total surface area of the entropic tetrahedron fixed by $S$, given by $F_{4,\text{max}}^{(S)}=S/4$. Applying this to the two highly entangled states, we have $\mathcal{P}(\text{GHZ})=1$ and $\mathcal{P}(\phi_4)=0.976$.

As a further illustration, we examine the generalized GHZ states represented by $\cos{\theta}|0000\rangle+\sin\theta|1111\rangle$. We emphasize that, despite having varying $F_4$ values denoting their entanglement, these states consistently exhibit a maximal degree of permutation symmetry, $\mathcal{P}=1$, for all possible angles of $\theta$. This observation serves as a validation of the suitability for our defined quantity $\mathcal{P}$, as these states inherently exhibit permutation invariance among all four qubits.

{\it Scalability}.---Continuing along the geometric journey, the generalization to systems with five or more qubits presents an intriguing direction for future exploration, involving hypervolume of simplices in dimensions larger than three, making it challenging to obtain an intuitive visual representation. 

Nonetheless, our tetrahedron measure presented in this work maintains its scalability. Specifically, we can consistently partition a large system into four groups and explore genuine quadripartite entanglement among these groups. Essentially, the tetrahedron measure is a genuine four-qudit entanglement measure, where each party possesses more than two dimensions. The extension of entropic fill to qudit systems has been made possible by a recent study \cite{yang2022}. Additionally, we have conducted numerical verifications, demonstrating the existence of unique nonnegative solutions for Eq.~\eqref{eqn1} and \eqref{eqn2} in qudit systems up to $d=50$. Consequently, we conjecture that our tetrahedron measure is also applicable to qudit systems.

An immediate application of this result is the investigation of quantum information scrambling, a phenomenon that describes the redistribution and transformation of information within quantum many-body systems. However, previous studies of information scrambling only involved entanglement across bipartitions, overlooking the rich landscape of multipartite entanglement \cite{lewis2019unifying,choi2020quantum,zhu2022observation}. This omission comes from the absence of a suitable measure to quantify GME.

With our novel GME measure relying on symmetric properties of quantum states, we can explore various GME properties in many-body systems, both theoretically and experimentally. This can involve studying the ``velocity'' of GME generation, enhancing our comprehension of distinct timescales within early stages of thermalization following quantum quenches. We also anticipate the identification of new phases, such as the ``GME volume law'' and the ``GME area law.''

{\it Mixed-state extension}.---The convex-roof construction generalizes entropic fill to mixed states $\rho$ as $F_4(\rho)=\min\sum_ip_iF_4(\psi_i)$, with the minimization taken over all pure-state realizations $\rho=\sum_ip_i|\psi_i\rangle\langle\psi_i|$. Although this method is computationally intensive, numerical techniques were developed to evaluate this quantity \cite{rothlisberger2009,eisert2007,xie2022esd}. Recently, a method was proposed to estimate convex-roof based entanglement measures in experiments \cite{xie2023experimental}.

The authors thank Professors Foek Hioe, Sabre Kais, Gabriel T.~Landi, Peter W.~Milonni, and Akira Sone for valuable discussions. Financial support was provided by National Science Foundation Grants No.~PHY-1505189 and No.~PHY-1539859 (INSPIRE), and a competitive grant from the University of Rochester.

\bibliography{tetrahedron}

\end{document}